\documentclass[a4paper, 11pt]{article}

\def\b{\begin{eqnarray}}
\def\e{\end{eqnarray}}
\def\n{\noindent}

\begin{document}

\begin{center}

{\LARGE\textbf{Pseudo-Newtonian Potential for Charged \vskip .1cm Particle in Kerr--Newman Geometry \\}} \vspace {10mm}
\vspace{1mm}
\noindent
{\large \bf Rossen I. Ivanov$^\dagger$ and \setcounter{footnote}{6}
Emil M. Prodanov$^\ast$}
\vskip1cm
\n
\hskip-.3cm
\begin{tabular}{c}
\hskip-1cm
$\phantom{R^R}^\dagger${\it Department of Pure and Applied Mathematics, Trinity College,}
\\ {\it University of Dublin, Ireland, and} \\
{\it Institute for Nuclear Research and Nuclear Energy, Bulgarian} \\ {\it Academy of Sciences,
72 Tsarigradsko chaussee, Sofia -- 1784, Bulgaria} \\ {\it e-mail: ivanovr@tcd.ie} \\
\\
\hskip-.8cm
$\phantom{R^R}^\ast${\it School of Physical Sciences, Dublin City University, Glasnevin, Ireland} \\
{\it e-mail: prodanov@physics.dcu.ie}
\end{tabular}
\vskip1cm
\end{center}

\vskip2cm

\begin{abstract}
\n
We consider the equatorial circular motion of a test particle of specific charge $q/m \ll 1$  in the Kerr-Newman
geometry of a rotating charged black hole. We find the particle's conserved energy and conserved projection of
the angular momentum on the black hole's axis of rotation as corrections, in leading order of $q/m$, to the
corresponding energy and angular momentum of a neutral particle. We determine the centripetal force acting on
the test particle and, consequently, we find a classical pseudo-Newtonian potential with which one can mimic
this general relativistic problem.
\end{abstract}

\newpage

\n
In 1924, Manev \cite{man} (see also \cite{hag}) introduced a classical potential in order to modify the celestial
mechanics in accordance with the general-relativistic description. To describe the motion of a particle of mass
$m$ in the static field of universal gravitation, due to mass $M$, Manev replaced the mass $m$ with $m = m_0
\mbox{ exp}(M/r)$, where $m_0$ is an invariant and geometrized units $G = 1 = c$ are used. This led to the following
modification of Newton's gravitational law:
\b
F = \frac{M m_0}{r^2} \left(1 + \frac{3 M}{r} \right) \, .
\e
Potential of the type:
\b
\label{newt}
V(r) = A \frac{M}{r} + B \frac{M^2}{r^2} \, ,
\e
where $A$ and $B$ are some real constants and $M$ is the mass of the central body, was originally introduced
by Newton himself in his Philosophi\ae \, Naturalis Principia Mathematica (Book I, Article IX, Proposition XLIV,
Theorem XIV, Corollary II), in order to describe deviations in the Moon's orbit from the Keplerian laws. However,
modification of Newtonian physics for mimicking general-relativistic results is attributed to Manev. We will
refer to potential of the type (\ref{newt}) as to Manev potential. \\
Modeling the particle motion in the most general stationary asymptotically-flat space-time --- that of
Kerr and Newman \cite{kerr, newman, mtw} --- is a fundamental problem with many astrophysical applications.
This area has been under intense investigation and various pseudo-Newtonian potentials for Kerr \cite{kerr, mtw}
and Schwarzschild \cite{sch, mtw} geometries have been proposed \cite{other}: interesting physical applications,
for example in satelite theory, have been studied in \cite{phys}, different mathematical aspects in \cite{math}.
In this paper we model the equatorial circular motion of a lightly charged particle ($q/m \ll 1$) in
Kerr--Newman geometry by introducing a classical pseudo-Newtonian potential. An immediate area of application of a
classical pseudo-Newtonian effective potential of a charged particle is the description of accretion disks of
charged particles around a rotating charged massive centre. In 1989, De R\'ujula, Glashow and Sarid \cite{glashow}
proposed the idea that dark matter in our Universe is made of charged massive particles rather than neutral particles.
These authors studied a galactic halo of such charged massive particles together with its cosmological implications,
including galactogenesis. Gould {\it et al.} \cite {gould} have shown that if charged massive particles (with masses
bewteen $10^2$ GeV  and  $10^{16}$ GeV) made up the dark halo of the Galaxy, then they would be present in large
numbers in disk stars. Charged particles and their effects on galaxies, stars and planets have been studied by
Dimopoulos {\it et al.} \cite{dimo}.

\newpage
\n
Rotating charged black holes are described by the Kerr--Newman geometry \cite{kerr, newman}. The Kerr--Newman metric in
Boyer--Lindquist coordinates \cite{bl} (see also \cite{mtw}), in geometrized units $G = 1 = c$, is given by:
\b
ds^2 & = & - \, \, \frac{\Delta}{\rho^2} (dt - a \sin^2 \theta \,\, d\phi)^2 +
\frac{\sin^2 \theta}{\rho^2} \Bigl[ a \, dt  - (r^2 + a^2) d\phi \Bigr]^2 \nonumber \\
& & \hskip6cm + \, \, \frac{\rho^2}{\Delta} dr^2 + \rho^2 d \theta^2 \, ,
\e
where
\b
\Delta & = & r^2 - 2 M r + a^2 + Q^2\, , \\
\rho^2 & = & r^2 + a^2 \cos^2 \theta \, .
\e
In the above, $M$ is the mass of the centre, $a > 0$ --- the specific angular momentum of the centre
(i.e. angular momentum per unit mass) and $Q$ --- the charge of the centre. \\
The motion of a particle of mass $m$ and charge $q$ in gravitational and electromagnetic fields is governed
by the Lagrangian:
\b
L = \frac{1}{2} \, \, g_{ij} \frac{d x^i}{d \lambda} \frac{d x^j}{d \lambda} - \frac{q}{m} \, A_i \, \frac{dx^i}{d\lambda} \, .
\e
In the above, $\lambda$ is the proper time $\tau$ per unit mass $m$: $\lambda = \tau/m$ and $A$ is the vector electromagnetic
potential, determined by the black hole's charge $Q$ and specific angular momentum $a$:
\b
A_i dx^i = - \frac{Q r}{\rho^2} (dt - a \, \sin^2 \theta \, d \phi) \, .
\e
(The magnetic field is due to the dragging of the inertial reference frames into rotation around the black hole.) \\
The equations of motion for the particle are:
\b
\label{geo}
\frac{d^2 x^i}{d \tau^2} + \Gamma^i_{jk} \, \frac{dx^j}{d \tau} \, \frac{dx^k}{d \tau} =
\frac{q}{m} F^i_{\phantom{ii}j} \, \frac{d x^j}{d \tau} \, ,
\e
where $F = dA$ is Maxwell's electromagnetic tensor and $\Gamma^i_{jk}$ are the Christoffel symbols. \\
For Kerr--Newman geometry, the geodesic equations (\ref{geo}) can be written as \cite{car} (see also \cite{fn}):
\b
\rho^2 \, \frac{dt}{d \lambda} &  =  & - a^2 E \sin^2 \theta + a J
    + \frac{r^2 + a^2}{\Delta} \Bigl[ E(r^2 + a^2) - J a - q Q r \Bigr] \, ,  \\ \nonumber \\
\label{potential}
\rho^2 \, \frac{dr}{d \lambda} & = & \pm \sqrt{\Bigl[ E(r^2 + a^2) - J a - q Q r \Bigr]^2
    - \Delta \Bigl[ m^2 r^2 + (J - a E)^2 + K \Bigr]} \, , \nonumber \\ \\
\rho^2 \, \frac{d\theta}{d \lambda} & =  & \pm \sqrt{K - \cos^2 \theta \Bigl[ a^2(m^2 - E^2)
    + \frac{1}{\sin^2 \theta} \, J^2 \Bigr]} \, , \\ \nonumber \\
\rho^2 \, \frac{d\phi}{d \lambda} &  =  & - a E + \frac{J}{\sin^2 \theta}
    + \frac{a}{\Delta} \Bigl[ E(r^2 + a^2) - J a - q Q r \Bigr] \, ,
\e
where $E = \partial L / \partial \dot{t}$ is the conserved energy of the particle,
$J = \partial L / \partial \dot{\phi}$ is the conserved projection of the particle's angular momentum on the axis
of the black hole's rotation (dots denote differentiation with respect to $\lambda$).  $K$ is another conserved
quantity given by:
\b
K = p_\theta^2 + \cos^2 \theta \Bigl[ a^2(m^2 - E^2)+ \frac{1}{\sin^2 \theta} \, J^2 \Bigr].
\e
Here $p_\theta$ is the $\theta$-component of the particle's four-momentum:
$p_\theta = \partial L / \partial\dot{\theta}$.  \\
The radial equatorial ($\theta = \pi/2 \, $) motion of the particle can be modelled as a one-dimensional motion
of a classical particle in effective potential $V(r)$ determined from (\ref{potential}):
\b
\frac{1}{2}\Bigl(\frac{dr}{d \lambda}\Bigr)^2 & \!\!\!\!\! = \!\!\!\! & V(r) \nonumber \\
& \!\!\!\!\! = \!\!\!\! & \frac{1}{2 \rho^4} \biggl[ \Bigl[ E(r^2 + a^2) - J a - q Q r \Bigr]^2
\! - \! \Delta \Bigl[ m^2 r^2 + (J - a E)^2 \Bigr] \biggr].
\e
We now focus on circular orbits in the equatorial plane. For circular orbits, $dr/d\lambda = 0$, both instantaneously
and at all subsequent times (orbit at a perpetual turning point) \cite{bardeen, muh}. This implies:
\b
\label{uno}
V(r) & = & 0 \, , \\
\nonumber \\
\label{due}
\frac{d V(r)}{d r} & = & 0 \, .
\e
Equations (\ref{uno}) and (\ref{due}) can be viewed as a system of simultaneous equations from which one can
determine the conserved energy $E$ of the particle and the conserved projection $J$ of the particle's angular
momentum on the axis of the black hole's rotation in terms of the parameters $M$, $Q$, $a$, $q$, and $m$. These
equations are non-linear and we will not be looking for their general solution. Instead, we will assume that the
particle's specific charge is small, i.e. $q/m << 1$, and we will expand $E$ and $J$, as functions of $q/m$,
with $m$ fixed, in Taylor series near $q/m = 0$:
\b
\label{expE}
E \Bigl(\frac{q}{m}\Bigr) & \!\!\!\! =  \!\!\!\!
& E(0) + E'(0) \Bigl( \frac{q}{m}\Bigr) + O \Bigl[ \Bigl( \frac{q}{m}\Bigr)^2 \Bigr] \, \equiv
    \, E_0 + q E_1 + O \Bigl[ \Bigl( \frac{q}{m}\Bigr)^2 \Bigr] , \\
\label{expJ}
J \Bigl(\frac{q}{m}\Bigr) &  \!\!\!\!  =  \!\!\!\!
& J(0) + J'(0) \Bigl( \frac{q}{m}\Bigr) + O \Bigl[ \Bigl( \frac{q}{m}\Bigr)^2 \Bigr] \, \equiv
    \, J_0 + q J_1 + O \Bigl[ \Bigl( \frac{q}{m}\Bigr)^2 \Bigr] .
\e
In the above expansions, one can identify $E(0) \equiv E_0$ as the conserved energy and $J(0) \equiv J_0$ as the
conserved projection of the angular momentum of a neutral particle in Kerr--Newman Geometry. These are known
\cite{dadhich}:
\b
\label{E0}
\frac{E_0}{m} & =  & \frac{r^2 - 2 M r + Q^2 \pm \, a \sqrt{Mr - Q^2}}
    {r \sqrt{r^2 - 3 M r + 2 Q^2 \pm 2 a \sqrt{M r - Q^2}}}, \nonumber \\ \\
\label{J0}
\frac{J_0}{m} & = & \frac{ \pm \, a (Q^2 - 2 M r) + (a^2 + r^2) \sqrt{M r - Q^2}}
        {r \sqrt{r^2 - 3 M r + 2 Q^2 \pm 2 a \sqrt{M r - Q^2}}} \, .
\e
The upper sign corresponds to direct orbit ($J>0$), while the lower sign corresponds to retrograde orbit ($J<0$). \\
The above expressions are valid for $r > Q^2/M$. In addition, the condition $M^2 \ge a^2 + Q^2$ is necessary for the
existence of a horizon (i.e. a black hole solution) in Kerr--Newman geometry \cite{mtw}. Orbits do not exist for
all values of $r > Q^2/M$. The question of existence and stability of bound circular orbits of neutral particles
in Kerr--Newman geometry, as well as the thresholds for photon, marginally stable and marginally bound orbits are
analysed in \cite{dadhich} (see also \cite{bardeen} for rotating black holes). Our goal is to find a classical
pseudo-Newtonian potential for modelling the circular orbital motion of a lightly charged particle ($q/m \ll 1$)
in Kerr--Newman geometry by seeking expansions for large values of $r$ (i.e. far from the black hole) which are
above the threshold of a marginally stable orbit, i.e. within our approximation, all orbits are stable.  \\
Identifying the Keplerian angular momentum distribution $J/E$, the centripetal force $F_0$ acting on the neutral
particle can be expressed as \cite{muh}:
\b
\label{f}
F_0 & = & \frac{J_0^2}{E_0^2 r^3} \nonumber \\
& = & \, r^{-3}  \biggl[a^3 (Q^2 - Mr)  + a (Q^4 + 4M^2r^2 - 3 Mr^3 - 4Q^2 Mr + 2 Q^2 r^2) \biggr. \nonumber \\
&& \left. \pm \, \,  r^2 \Delta \sqrt{M r - Q^2} \,\, \right]^2
\left[a^2 (Q^2 - Mr) - (Q^2 - 2Mr + r^2)^2 \right]^{-2}.
\e
Returning to the case of charged particle, we substitute the expansions (\ref{expE}) and (\ref{expJ}) into equations
(\ref{uno}) and (\ref{due}) and we discard terms quadratic in $q/m$. We also use the fact that (\ref{E0}) and (\ref{J0})
are solutions to equations (\ref{uno}) and (\ref{due}) with $q/m = 0$. Thus, equations (\ref{uno}) and (\ref{due})
become:
\b
\label{edno1}
& & \hskip-1cm (r^2 + a^2) E_1 - a J_1 - \Delta \frac{J_0 - a E_0}{(r^2 + a^2) E_0 - a J_0} (J_1 - a E_1)
    \: = \: Q r \, \\
\label{dve2}
& & \hskip-1cm 2 (M - r) J_0 J_1 - 2 a M (J_0 E_1 + E_0 J_1) + 2 (a^2 M + a^2 r + 2 r^3) E_0 E_1 \nonumber \\
& & \hskip5cm = \: - a Q J_0 + Q (a^2 + 3 r^2) E_0 \, .
\e
These are now linear equations and they can be easily solved:
\b
\label{E1}
E_1 & \!\!\!\! \equiv  \!\!\!\! & \frac{E'(0)}{m} = \frac{Q \Bigr[ \pm \, 4 a (M r - Q^2) + (r^2 - 4 M r - a^2 + 3Q^2)
    \sqrt{M r - Q^2} \Bigr]}
    {2 r \Bigr[ \pm \, 2 a (M r - Q^2) + (r^2 - 3 M r + 2 Q^2) \sqrt{M r - Q^2} \Bigr]}, \nonumber \\ \\
\label{J1}
J_1 & \!\!\!\! \equiv \!\!\!\! & \frac{J'(0)}{m} = \pm \, \frac{Q \Bigl[ \mp \, a (r^2 - 4 M r + 3 Q^2 + a^2)
    \sqrt{M r - Q^2} \Bigr]}
    {2 r \Bigr[ \pm \, 2 a (M r - Q^2) + (r^2 - 3 M r + 2 Q^2) \sqrt{M r - Q^2} \Bigr]} \nonumber \\
& & \hskip1.45cm \pm \frac{Q \Bigl[(a^2 + r^2) (r^2 - 4 M r + Q^2) + 2 M r^3 + 3 a^2 Q^2\Bigr]}
    {2 r \Bigr[ \pm \, 2 a (M r - Q^2) + (r^2 - 3 M r + 2 Q^2) \sqrt{M r - Q^2} \Bigr]}\!.
\e
As before, upper sign corresponds to direct orbit ($J>0$), while lower sign corresponds to retrograde orbit
($J<0$). \\
The centripetal force acting on the charged particle is:
\b
F  \, = \,  \frac{1}{r^3} \, \frac{J^2}{E^2} & = & \frac{1}{r^3} \biggl( \frac{J_0}{E_0} \biggr)^2
\Biggl( \frac{1 + \frac{q}{m}\frac{J_1}{J_0}}{1 + \frac{q}{m} \frac{E_1}{E_0}} \Biggr)^2  \nonumber \\
& = & F_0 \biggr[ 1 + 2 \frac{q}{m} \Bigl( \frac{J_1}{J_0} - \frac{E_1}{E_0} \Bigr) +
O \Bigl[ \Bigl( \frac{q}{m} \Bigr)^2 \Bigr] \biggl].
\e
The pseudo-Newtonian potential $V_M(r)$ giving rise to this force is the Manev potential,
\b
\label{manev}
V_M(r) = - \int F dr \, ,
\e
which, upon expansion over the powers of $1/r$, results in:
\b
V_M(r) & = & \Bigl( 1 - \frac{q}{m} \frac{Q}{M} \Bigr) \: \frac{M}{r}
    \:\: + \:\: \Bigl( 2 - \frac{1}{2} \, \frac{Q^2}{M^2} - \frac{9}{4} \,  \frac{q}{m} \frac{Q}{M} \Bigr)
    \: \frac{M^2}{r^2}  \nonumber \\ \nonumber \\
& & \mp \:\: \frac{12}{5} \Bigl( 1 - \frac{q}{m} \frac{Q}{M}\Bigr) \: \frac{a M^{3/2}}{r^{5/2}}
    \nonumber \\ \nonumber \\
& & + \Bigl[ 4 + \frac{2}{3} \frac{a^2}{M^2} - 2 \frac{Q^2}{M^2} + \frac{q}{m} \Bigl( - \frac{37}{8} \frac{Q}{M}
                    - \frac{2}{3}\frac{a^2 Q}{M^3} + \frac{Q^3}{M^3} \Bigr)\Bigr] \frac{M^3}{r^3}
    \nonumber \\ \nonumber \\
& & \mp \Bigl[ 8  - 2 \frac{Q^2}{M^2} + \frac{q}{m} \Bigl( - \frac{60}{7} \frac{Q}{M}
        + \frac{4}{7} \frac{Q^3}{M^3} \Bigr) \Bigr] \frac{a M^{5/2}}{r^{7/2}} \: \nonumber \\ \nonumber \\
& & + \Bigl[ 8 +  \frac{23}{4} \frac{a^2}{M^2} - 6 \frac{Q^2}{M^2}  - \frac{1}{2} \frac{a^2 Q^2}{M^4}
              + \frac{1}{2} \frac{Q^4}{M^4} \nonumber \\ \nonumber \\
& & \hskip2cm
        + \frac{q}{m} \Bigl( - \frac{569}{64} \frac{Q}{M}
            - \frac{47}{8} \frac{a^2 Q}{M^3} + \frac{33}{8} \frac{Q^3}{M^3} \Bigr) \Bigr] \: \frac{M^4}{r^4} \nonumber \\
& & + \cdots
\e
From this expression it is evident that the leading term is reduced due to Coulomb repulsion if the two charges
$q$ and $Q$ have the same sign. Opposite charges lead to additional attraction. By setting $a=0$ one obtains the
effective pseudo--Newtonian potential for Reissner--Nordstr\o m geometry \cite{rn}, and by further setting
$Q = 0$ --- for Schwarzschild geometry \cite{sch} (see also \cite{mtw}).


\end{document}